\documentstyle[psfig,mnras_cite,times]{mn}
\begin{document}
\author[Elspeth M. Minty, Alan F. Heavens, Michael R.S. Hawkins]
{Elspeth M. Minty, Alan F. Heavens, Michael R.S. Hawkins\\
Institute for Astronomy, University of Edinburgh, Royal
Observatory, Blackford Hill, Edinburgh EH9 3HJ , United Kingdom}
\date{\today}
\title{Testing dark matter with high-redshift supernovae}
\newcommand{\be}{\begin{equation}}
\newcommand{\ee}{\end{equation}}
\newcommand{\ba}{\begin{eqnarray}}
\newcommand{\ea}{\end{eqnarray}}
\def\gs{\mathrel{\raise1.16pt\hbox{$>$}\kern-7.0pt 
\lower3.06pt\hbox{{$\scriptstyle \sim$}}}}         
\def\ls{\mathrel{\raise1.16pt\hbox{$<$}\kern-7.0pt 
\lower3.06pt\hbox{{$\scriptstyle \sim$}}}}         

\maketitle

\begin{abstract}
Dark matter in the Universe consisting of macroscopic objects such as
primordial black holes may cause gravitational lensing of distant
objects.  The magnification associated with lensing will lead to
additional scatter in the received flux from standard candles, and too
small an observed scatter could rule out compact dark matter entirely.
In this letter, we show how the scatter in fluxes of distant Type 1a
supernovae could be used to distinguish between models with and
without lensing by macroscopic dark matter.  The proposed SNAP
project, with $\sim 2400$ supernovae in the range $0.1\ls
z\ls 1.7$, should be able to identify models at $99.9$\% confidence,
if systematic errors are controlled.  Note that this test is
independent of any evolution of the mean supernova luminosity with
redshift.  The variances of the current Supernova Cosmology Project
sample do not rule out compact lenses as dark matter: formally they
favour such a population, but the significance is low, and removal of
a single faint supernova from the sample reverses the conclusion.
\end{abstract}

\section{Introduction}

In recent years one of the classic tests of the geometry of the
Universe has undergone a resurgence of interest.  Type 1a
supernovae are thought to act as standard candles, and therefore
their Hubble diagram may be used to constrain cosmological
models.  Dedicated discovery and follow-up programmes
(\pcite{Perl97}, \pcite{Perl98}, \pcite{Schmidt98}) have
established the Hubble diagram to $z \simeq 1$, leading to
constraints on $H_0$, $\Omega_m$ and $\Omega_\Lambda$, most
notably the requirement of a positive cosmological constant
\cite{Riess98}.  It has long been recognised that the supernova Hubble
diagram would be affected by gravitational lensing, if significant
quantities of dark matter resided in the form of macroscopic compact
objects (herafter MACHOs) such as black holes (\pcite{LSW88}, 
\pcite{Rauch91}, \pcite{Holz98},
\pcite{MS99}, \pcite{Perl99}, \pcite{SH99}, \pcite{WA2000},
\pcite{HT2000}).  
Such a population has been proposed to account for the long-term
variability of quasars \cite{Hawkins93}.  

These studies have shown that lensing can have a significant effect on
the estimation of cosmological parameters, and that the distribution
of supernova fluxes could be used to determine whether the dark matter
was in MACHO form or not.  \scite{MS99} showed that, provided the
cosmology was known, microscopic dark matter could be distinguished
from MACHOs with relatively small numbers of supernovae.  The reason
for this sensitivity is that in a MACHO-dominated universe of
reasonable density, most bunches of light rays do not undergo large
magnifications, and the most probable flux received is close to that
expected in empty-beam models \cite{DyerRoeder74}.  This shifts the
most probable Hubble diagram systematically, changing the estimates of
the cosmological parameters \cite{Holz98}.  Given that one does not
know the underlying cosmology, one should estimate simultaneously the
background cosmology as well as the contribution to the matter density
by MACHOs.  This is ambitious, and will surely be attempted once
larger supernova searches such as those proposed by the
Supernova/Acceleration Probe (SNAP: see http://snap.lbl.gov) and by
the Visible and Infrared Telescope for Astronomy (VISTA;
http://www.vista.ac.uk) are underway.  In this paper, the study is
more limited; we focus on how the extra scatter in supernova fluxes,
from lensing, can be used to constrain the quantity of MACHO dark
matter. The variance will have contributions from intrinsic variations
in supernova properties, instrumental error, and lensing.  The first
two of these should be virtually independent of the cosmological
model; the presence or absence of significant additional scatter
allows a test of whether the dark matter is in the form of MACHOs.

The detectability of dark matter candidates in this test rests on how
accurately the variance can be estimated from sets of supernovae.  For
gaussian distributions, this is readily done, but, although it may be
a reasonable approximation to model the intrinsic and instrumental
effects as gaussian, the lensing effect is highly skewed towards rare
high-magnification events.  We must therefore take care in modelling
this effect accurately.  We use numerical ray-tracing simulations
(described in section 2) and analyse the results using a Bayesian
method (section 2.1).  Lensing induces an extra scatter which rises
with redshift, contributing as much as $0.5$ magnitudes at a redshift
1.5 for an Einstein-de Sitter model with all the matter in MACHOs.
This is readily distinguishable from a no-lensing model, provided
sufficient supernovae are available.  We find, in general, that with
$\sim 2400$ supernovae selected from $0.1 \ls z
\ls 1.7$, as proposed by SNAP, the scatter alone can select
between several cosmological models at $99.9$\% confidence level, if
systematic errors are controlled.  Note that this test has the
advantage that it is insensitive to any evolution in the mean
luminosity of supernova (although it is dependent on any evolution in
the scatter of intrinsic properties).  Similar conclusions to ours
have recently been reached via a slightly different analysis by
\cite{MGB01}.

\section{Method}

\begin{figure}
\centerline{
\psfig{figure=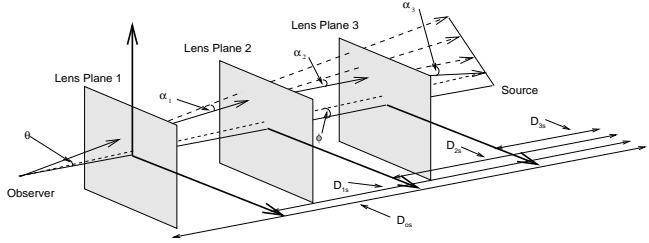,width=8.5cm,angle=0}}
\caption[]{\label{raytracing} The ray-tracing simulation geometry. In
practice we use 5 planes, but test with more.}
\end{figure}

\begin{figure}
\centerline{
\psfig{figure=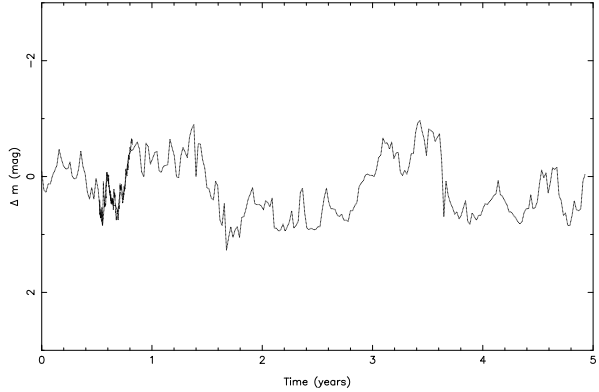,width=8.5cm,angle=270}}
\caption[]{\label{lightcurve} A sample simulated lightcurve of a
supernova at $z=1.7$ in an Einstein-de Sitter universe dominated by
MACHOs.  The timescale is set by the mass of lenses and their velocity
dispersion; for the purposes of this figure, these are
$10^{-4}M_\odot$ and $300$ km s$^{-1}$.}
\end{figure}

The microlensing effect on the brightness of supernovae at redshifts
$\sim 1$ is not necessarily accurately modelled by a single-scattering
event.  We therefore simulate the lensing of distant supernovae with a
ray-tracing program.  We assume the Universe is populated with compact
dark matter candidates of a single mass.  The results are independent
of the mass, provided the source is small compared to the Einstein
radius (for canonical expansion speeds, at maximum light this is $\sim
10^{13} m$, which requires $M\gs 10^{-2} M_\odot$).  The masses are
confined to a discrete set of lens planes, and rays are passed from
the observer to the source plane, being bent at each stage by the
lenses in each plane.  The bend angle is the vector sum of the bends
from individual lenses, and is computed using the Barnes-Hut tree code
\cite{Barnes86}.  The bend angle for each lens is directed towards each lens,
with magnitude $\alpha = {4GM\over r c^2}$ where $r$ is the impact
parameter.  The rays are traced to the source redshift, and the solid
angle subtended by each of the distorted pixels is computed. Since
surface brightness is conserved by the lensing process, the ratio of
the solid angles of the image plane pixel and distorted source plane
pixel gives the magnification.  Full details of the simulations will
appear elsewhere (Minty et al, in preparation), but in summary we use
$400-5000$ lenses in total, randomly placed on $5$ planes located to
give equal numbers of lenses per plane.  The random hypothesis is
justified in \scite{HW98} and \scite{Met99}. Backwards rays are sent
out on a $1024 \times 1024$ grid (see Fig. \ref{raytracing}), with
rays separated by a small fraction of the Einstein radius of lenses on
the nearest plane, ensuring the pattern is fully sampled.  In
computing the bend angle, the lens pattern is assumed to be periodic.
We perform tests varying the number of sources, size of lens planes,
number of planes, tree-code parameters, grid size, to ensure
robustness of the results.  The need for numerical simulations is
illustrated in fig. \ref{lightcurve}, which shows a typical lightcurve 
for lensing in an Einstein-de Sitter Universe, where the dark matter
is all in lenses.  The optical depth to lensing is high, and the
magnification is not well described by a single-scattering event.
We consider four separate models, one with no lensing, one an
Einstein-de Sitter model with all matter in lenses, and two flat
models with matter $\Omega_m=0.3$, and cosmological constant
$\Omega_\Lambda=0.7$.  These two models differ in the proportion of
the matter in lenses.  The models are detailed in table 1.  We perform
two analyses here, a preliminary study of the 42 high-redshift
supernovae in the Supernova Cosmology Project \cite{Perl99}, and a
second study investigating whether SNAP should be capable of detecting
MACHOs unambiguously.  In the former case, we take the errors from the
obsevations.  For the SNAP study, we assume that there is an intrinsic
variation in supernovae properties of 0.157 mag
\cite{Perl99}, and a measurement error of 0.08 mag \cite{MS99}.
These errors (in magnitude) are assumed to be gaussian, and
independent of redshift and cosmology, although these assumptions
could be relaxed if desired.  Model 1 contains only these components
of variance.
\begin{table}
\begin{tabular}{l|l|l|l}
\hline
Model & Matter density & Cosmological constant & Density in lenses \\
      &  $\Omega_m$ & $\Omega_\Lambda$ & $\Omega_{Lens}$ \\
\hline
1     &   & & 0 \\
2     & 0.3 & 0.7 & 0.1 \\
3     & 0.3 & 0.7 & 0.3 \\
4     & 1.0 & 0.0 & 1.0 \\
\hline
\end{tabular}
\caption[]{\label{models} Models considered.  Model 1 includes scatter 
only from measurement error and intrinsic variations.}
\end{table}
\begin{figure}
\centerline{
\psfig{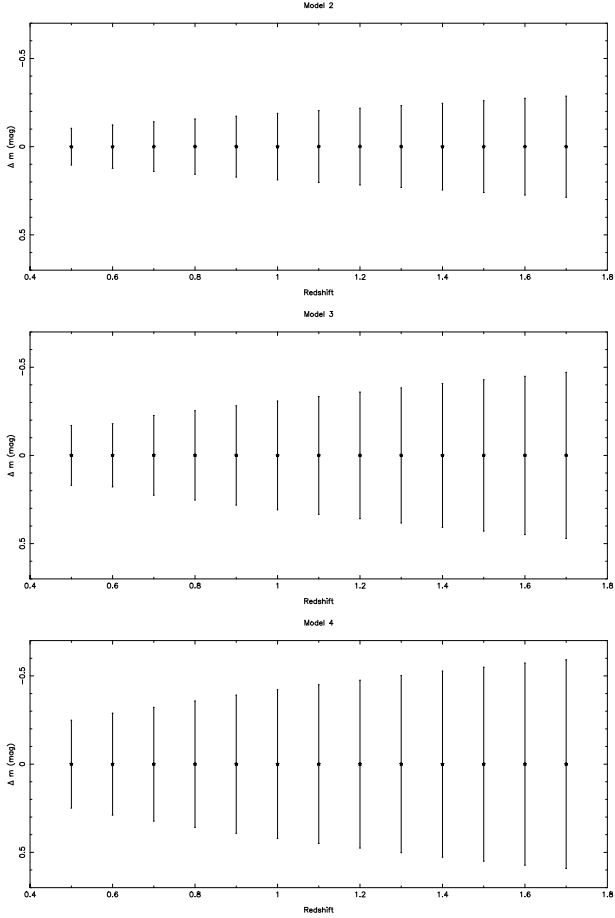}}
\caption[]{\label{scatter} Mean and r.m.s. deviation of supernova
brightness vs. redshift, due to lensing alone.  Note that the scatter
has been symmetrised.  The models have, from the top,
($\Omega_0,\Omega_\Lambda,\Omega_{Lens}$)=(0.3,0.7,0.1), (0.3,0.7,0.3) 
and (1.0,0.0,1.0).}
\end{figure}
Fig. \ref{scatter} shows the scatter in the magnitudes induced by
lensing for models 2--4, as a function of redshift.  Given that the
intrinsic plus instrumental scatter is $\sim 0.2$ mag, we see that the additional
variance from lensing becomes very significant at redshifts $\gs 1$.

\subsection{Statistics}

\begin{figure}
\psfig{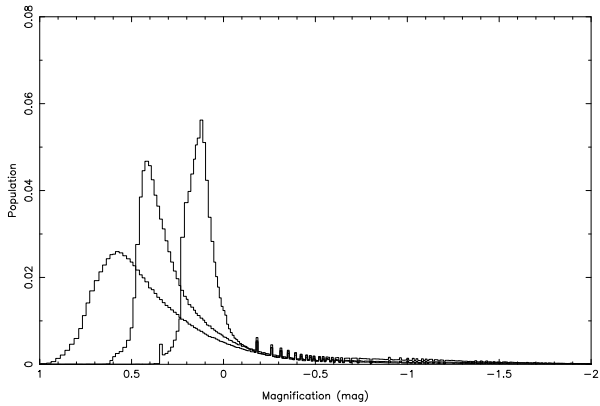}
\caption[]{\label{magdist} The distribution of magnifications arising
from lensing in the three lensing models at redshift $z=1.7$. Models
2-4, with increasing $\Omega_{Lens}$, from the left.}
\end{figure}

\begin{figure}
\psfig{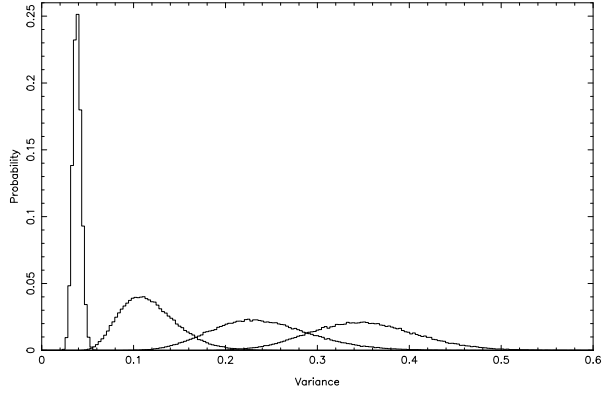}
\caption[]{\label{rmsdistribution}  The distribution of the
r.m.s. scatter of $N=150$ supernovae at redshift $z=1.7$, for the four
models (1--4, from the left).}
\end{figure}

SNAP proposes to obtain $N \sim 100-300$ supernovae per redshift
interval of $\Delta z=0.1$ between $z=0.4$ and $z=1.2$, with small
numbers at lower and higher redshift (up to $z \sim 1.7$; see the
Science Case in http://snap.lbl.gov for full details).  For
statistical analysis, we use the r.m.s. in supernova magnitude in
each redshift bin.  The discriminatory power is thus determined by how
accurately the r.m.s. can be estimated from $N$ supernovae.  This can
be calculated analytically for gaussian distributions, but the
distribution of magnifications induced by lensing is far from gaussian
(see fig. \ref{magdist}).
We therefore draw $N$ magnifications at random from the lensing
simulations, noting that the probability of a supernova lying in a
distorted pixel in the source plane is proportional to the solid angle
of the pixel.  These magnifications are applied to magnitudes drawn
from gaussian distributions of variance $\sigma^2 =
\sigma^2_{intrinsic} + \sigma^2_{observational}$.  For the
proposed SNAP survey we take $\sigma_{intrinsic}=0.157$ mag
\cite{Perl99} and $\sigma_{observational}=0.08$ mag \cite{MS99}.  For the current
data, we take $\sigma$ to be the individual estimated
r.m.s. \cite{Perl99}.  We do this repeatedly, and compute the
distribution of the sample variance as a function of redshift.  Some
representative distributions are shown in fig. \ref{rmsdistribution}.
We compute the probability distribution of the variance
estimator $D_i = \sum_j (m_{j,i}-\bar m_i)^2/(N-1)$, where the
sum extends over $j=1\ldots N$ supernovae in a redshift bin $i$.
$\bar m_i$ is the mean magnitude in bin $i$.   We do this by
Monte-Carlo simulation of $N$ supernovae drawn from ray-tracing
simulations.  This gives us, for a given model $M_k$, the
probability of obtaining a set of variances $\{D_i\}$,
\be
p(\{D_i\}|M_k,N) = \prod_i p(D_i|M_k,N)
\ee
since the bins are independent.  From now on, we suppress the $N$ in
these probabilities.

A complication arising for our analysis of future surveys is the
inevitable absence of data.  We seek the probability of deducing the
correct model $M_{cor}$.  Marginalising over the (unknown at this
stage) true model, this is
\ba
p(M_{cor}) &=& \sum_k p(M_{cor}, M_k) = \sum_k p(
M_{cor}|M_k) p(M_k)\\
&= &{1\over N_k}\sum_k p(M_{cor}|M_k)
\ea
where the last step follows if we assume equal prior
probabilities for the true model.  $N_k$ is the number of models
considered.

To compute the conditional probability in the last equation, we
use the distribution of sets of variances $\{D_i\}$, given that
the true model is $M_t$:
\be
p(M_{cor}|M_t) = \int d\{D_i\} p(M_{cor}|\{D_i\}))
p(\{D_i\}|M_t)
\label{pp}
\ee
Using Bayes' theorem,
\be
p(M_k|\{D_i\}) = {p(\{D_i\}|M_k)p(M_k)\over p(\{D_i\})}
\ee
we see that the probability of getting the correct model,
given a set of data is
\be
p(M_{cor}|\{D_i\}) = {p(\{D_i\}|M_t)p(M_t)\over \sum_k
p(\{D_i\}|M_k) p(M_k)}
\ee
where the evidence $p(\{D_i\})$ cancels out top and bottom.
If we assume uniform priors for the models, the probability
simplifies, and substitution into $(\ref{pp})$ gives
\be
p(M_{cor}|M_t) = \int {p(\{D_i\}|M_t)^2\over \sum_k
p(\{D_i\}|M_k) p(M_k)} d\{D_i\}
\ee
In practice we approximate the integral over sets of data by a set of
$N_{r}$ random drawings (labelled by $\alpha$):
\be
p(M_{cor}) = {1\over N_k}\sum_{M_t} {1\over N_{r}}
\sum_{\alpha} \left[{p(\{D_i\}_\alpha|M_t)\over \sum_k
p(\{D_i\}_\alpha|M_k) p(M_k)}\right].
\ee
We compute this probability for varying numbers of supernovae per unit 
redshift interval.

\section{Results}

For the existing 42 high-redshift supernovae published from the
Supernova Cosmology Project \cite{Perl99}, we can compute the relative
likelihood of the four models considered.  For the data, we use the
stretch luminosity-corrected effective B-band magnitude.  We subtract
from it the expected magnitude for model 2 (or 3), and compute the
variance (using the standard estimator) for 4 bins between $z=0.45$
and $z=0.85$.  The reason for the subtraction is that the redshift bins
for which we have microlensing amplification distributions are quite
broad, and the mean apparent magnitude is expected to vary
substantially over the bin.  For the Einstein-de Sitter model, the
additional variance from subtracting the wrong evolution is
negligible.
\begin{table}
\begin{tabular}{l|c|c}
\hline
Redshift & Number of supernovae & Variance \\
\hline
0.45-0.55     & 11 & 0.064 \\
0.55-0.65     & 8  & 0.170 \\
0.65-0.75     & 3  & 0.076 \\
0.75-0.85     & 3  & 0.043 \\
\hline
\end{tabular}
\caption[]{\label{results} Estimated variances in supernova apparent
magnitudes for four redshift bins.  Data from the Supernova Cosmology
Project \cite{Perl99}.}
\end{table}
The variances in the four bins are shown in table
\ref{results}. Various effects conspire to make the $0.55 < z < 0.65$
bin the crucial one: first, at lower redshift, the lensing makes
little difference to the variance; second, at higher redshift, there
are very few supernovae in the bins; finally, the observed variance is
high in the second bin.  Thus the second redshift bin favours models
with more microlensing, since the expected no-lensing variance is only
0.04.  Combining the results gives the relative likelihoods shown in
table
\ref{Likelihoods}, where the likelihood of the favoured model is set
to unity.  We see that the variances in the data have a slight
preference for a population of MACHOs, but the significance is low.
In fact removing a single supernova (SN1997K) from the dataset reduces
the variance to 0.053, reversing the conclusions and making the
no-lensing model the preferred choice (see table
\ref{Likelihoods}).  There is little a priori justification for
removing this point.  In fact it is anomalously faint, so it is not a good
candidate lensing event.  Thus, unsurprisingly, we are unable to learn 
much from this test on current data.

\begin{figure}
\psfig{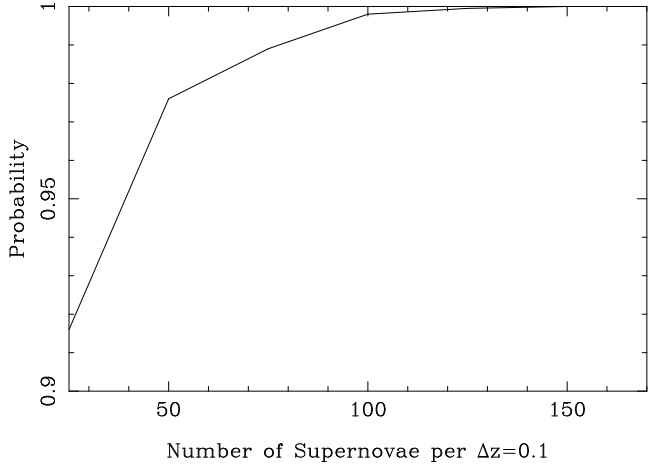}
\caption[]{\label{Pz} The probability of obtaining the correct
model, plotted against the number of supernovae per redshift interval
$\Delta z =0.1$.  In all cases it is assumed that supernovae are
observed up to $z=1.7$.}
\end{figure}

\begin{table}
\begin{tabular}{|l|l|l|l|l|l|}
\hline
Model &  $\Omega_m$ & $\Omega_\Lambda$ & $\Omega_{Lens}$ & 
Relative & SN1997K\\
& & & & likelihood & removed \\
\hline
1     &     &     & 0   & 0.58 & 1.0\\
2     & 0.3 & 0.7 & 0.1 & 0.92 & 0.88\\
3     & 0.3 & 0.7 & 0.3 & 0.96 & 0.66 \\
4     & 1.0 & 0.0 & 1.0 & 1.0  & 0.32\\
\hline
\end{tabular}
\caption[]{\label{Likelihoods} Relative likelihood for the four models,
from the high-redshift supernovae observed as part of the supernova
cosmology project.}
\end{table}
Future experiments should be able to do this task well.  In table
\ref{Probabilities},  we show the conditional probabilities of 
selecting models, given a correct model, for experiments
with a flat redshift distribution of supernovae in the range  $0.1 - 1.7$.
Probabilities less than 0.0005 are set to zero in the table.
\begin{table}
\begin{center}
\begin{tabular}{ccccccccc}
\hline
   & \multicolumn{7}{c}{Number of supernovae per $\Delta z = 0.1$} \\
      &  25   &  50   &  75   & 100   & 125   & 150   &  175  \\
\hline
\multicolumn{7}{l}{True Model: 1 ($\Omega_{Lens}=0$)}\\
1    & 0.998 & 1.000 & 1.000 & 1.000 & 1.000 & 1.000 & 1.000 \\
2    & 0.002 & 0.000 & 0.000 & 0.000 & 0.000 & 0.000 & 0.000 \\
3    & 0.000 & 0.000 & 0.000 & 0.000 & 0.000 & 0.000 & 0.000 \\
4    & 0.000 & 0.000 & 0.000 & 0.000 & 0.000 & 0.000 & 0.000 \\
\hline
\multicolumn{7}{l}{True Model: 2 ($\Omega_m=0.3,\ \Omega_\Lambda=0.7,\ 
\Omega_{Lens}=0.1$)}\\
1    & 0.004 & 0.000 & 0.000 & 0.000 & 0.000 & 0.000 & 0.000 \\
2    & 0.927 & 0.986 & 0.996 & 0.998 & 0.999 & 1.000 & 1.000 \\
3    & 0.069 & 0.014 & 0.004 & 0.002 & 0.001 & 0.000 & 0.000 \\
4    & 0.000 & 0.000 & 0.000 & 0.000 & 0.000 & 0.000 & 0.000 \\
\hline
\multicolumn{7}{l}{True Model: 3 ($\Omega_m=0.3,\ \Omega_\Lambda=0.7,\ 
\Omega_{Lens}=0.3$)}\\
1    & 0.000 & 0.000 & 0.000 & 0.000 & 0.000 & 0.000 & 0.000 \\
2    & 0.065 & 0.014 & 0.004 & 0.002 & 0.001 & 0.000 & 0.000 \\
3    & 0.838 & 0.949 & 0.977 & 0.990 & 0.996 & 0.999 & 1.000 \\
4    & 0.097 & 0.037 & 0.019 & 0.008 & 0.002 & 0.001 & 0.000 \\
\hline
\multicolumn{7}{l}{True Model: 4 ($\Omega_m=1.0,\ \Omega_\Lambda=0,\ 
\Omega_{Lens}=1.0$)}\\
1    & 0.000 & 0.000 & 0.000 & 0.000 & 0.000 & 0.000 & 0.000 \\
2    & 0.000 & 0.000 & 0.000 & 0.000 & 0.000 & 0.000 & 0.000 \\
3    & 0.088 & 0.032 & 0.016 & 0.006 & 0.003 & 0.001 & 0.000 \\
4    & 0.902 & 0.968 & 0.984 & 0.994 & 0.997 & 0.999 & 1.000 \\
\hline
\end{tabular}
\end{center}
\caption[]{\label{Probabilities} Probability of deducing, from the
four considered, correct and incorrect models given the true model.
Probabilities $<5 \times 10^{-4}$ are set to zero for this table.  The 
redshift distribution is assumed to be uniform across $0.1 < z < 1.7$.}
\end{table}
In fig. \ref{Pz} we show how the probability of obtaining the correct
model changes as we increase the number of sources per $\Delta z =
0.1$ bin.  We assume here a uniform prior: i.e. all models are equally
likely {\em a priori}.  SNAP's expected redshift distribution is not
uniform, but rises towards $z=1.2$ with a small number of supernovae
extending to $z=1.7$.  Using the expected redshift distribution from a
one-year SNAP experiment (see Science Case in http://snap.lbl.gov),
the conditional probabilities of selecting correct and incorrect
models are given in table \ref{SNAP}.  Basically SNAP should be able
to do this test very well; there is only a very small possibility of
confusing the $\Omega_{Lens}=0.3$ model with higher and lower lens
densities.  Combined with a uniform prior for the models, the
probability of selecting the correct model is 99.9\%.  In this paper,
we have only considered a discrete number of lensing models.  A full
Fisher matrix analysis of the error expected on the MACHO contribution 
is possible if many lensing simulations are run, or a good fitting
formula is used for the magnification probability, but it is clear
from this study that SNAP should be able to constrain the dark matter
nature extremely well.

\begin{table}
\begin{center}
\begin{tabular}{ccc}
\hline
True Model & Trial Model & p(Trial Model$|$True Model) \\ 
\hline
  1       &      1     &     1.0\\
          &      2     &     $2 \times 10^{-11}$\\
          &      3     &     0.0\\
          &      4     &     0.0\\
\hline
  2   &          1       &   0.0\\
      &          2       &   0.9999 \\
      &          3       &   $6 \times 10^{-5}$\\
      &          4       &   0.0\\
\hline
  3   &          1       &   0.0\\
      &          2       &   0.002\\
      &          3       &   0.994\\
      &          4       &   0.003\\
\hline
  4   &          1      &    0.0\\
      &          2      &    0.0\\
      &          3      &    0.0004 \\
      &          4      &    0.9996 \\
\hline
\end{tabular}
\end{center}
\caption[]{\label{SNAP}Probabilities of proposed 1-year SNAP mission
distinguishing between lensing and no-lensing models.  Model
parameters are given in table \ref{models}.}
\end{table}

\section{Conclusions}

We have demonstrated how the scatter in the supernova fluxes can be
used to support or rule out models with MACHO dark matter.  Lensing by
MACHOs increases the variance in the fluxes, and for models with a
high density in MACHO dark matter, the variance can be increased by a
factor greater than 3 at redshifts accessible by supernova searches.
Simply put, if the observed variance in supernova magnitudes is too
small, it can eliminate MACHOs as the dominant dark matter candidate.
The existing data from the Supernova Cosmology Project data do not
rule out MACHOs, and conclusions about whether MACHOs are preferred or
not are sensitive to inclusion or exclusion of individual supernovae.
Future planned surveys, such as the proposed SNAP survey should be able
to distinguish models readily, and in principle provide an accurate
measurement of the quantity of MACHO dark matter.

\noindent{\bf Acknowledgments}

We are grateful to Isobel Hook and Andy Taylor for useful discussions, and to 
the High Performance Computer Group, Hitachi Europe Ltd, on whose
machines some of the simulations were run.

\bibliographystyle{mnras}

\end{document}